\documentclass{ws-mpla}
\usepackage[super]{cite}
\usepackage{graphicx}
\usepackage{color}
\begin{document}

\markboth{H. B\"ohringer \&  G. Chon}
{Constraints on neutrino masses from a cosmological study of galaxy clusters}

\catchline{}{}{}{}{}


\title{Constraints on neutrino masses from the study of the
nearby large-scale structure and galaxy cluster counts}

\author{Hans B\"ohringer and Gayoung Chon}

\address{Max-Planck-Institut f\"ur extraterrestrische Physik,\\
                   D-85748 Garching, Germany\\
                   hxb@mpe.mpg.de}

\maketitle

\pub{Received (Day Month Year)}{Revised (Day Month Year)}

\begin{abstract}
The high precision measurements of the cosmic microwave 
background by the {\sf Planck} survey yielded tight constraints 
on cosmological parameters
and the statistics of the density fluctuations at the time
of recombination. This 
provides the means for a critical study of
structure formation in the Universe
by comparing the microwave background results with 
present epoch measurements of
the cosmic large-scale structure. It can reveal 
subtle effects such as how different forms of Dark Matter
may modify structure growth. Currently most interesting 
is the damping effect of structure growth
by massive neutrinos. Different observations of low redshift
matter density fluctuations provided evidence for
a signature of massive neutrinos. Here we discuss  
the study of the cosmic large-scale structure
with a complete sample of nearby, X-ray luminous clusters
from our {\sf REFLEX} cluster survey.
From the observed X-ray luminosity function and its reproduction
for different cosmological models, we obtain tight constraints 
on the cosmological parameters describing the matter density, $\Omega_m$,
and the density fluctuation amplitude, $\sigma_8$.
A comparison of these constraints with the {\sf Planck}
results shows a discrepancy in the framework of a pure $\Lambda$CDM
model, but the results can be reconciled, if we allow for a 
neutrino mass in the range of $0.17$ to $0.7$ eV. Also some 
others, but not all of the observations of the nearby 
large-scale structure provide
evidence or trends for signatures of  massive neutrinos.
With further improvement in the systematics 
and future survey projects, these indications will
develop into a definitive measurement of neutrino masses.

\end{abstract}


\section{Introduction}	
The recent high precision measurement of the cosmic microwave background
(CMB) by the {\sf Planck}
satellite has provided tight constraints on the parameters
of the cosmological model describing our Universe.
A $\Lambda$CDM model with six parameters gives a very good
description of the observations (Planck Collaboration 2015a).
Therefore the structure of our Universe
at the time of recombination (redshift of $\sim 1090$) 
is well defined. This opens the exciting 
possibility to study cosmic structure formation from the
recombination epoch
to the present and to look for very subtle effects, such as how
different types of ``Dark Matter'' modify the growth of structure.
The currently most interesting of these quests
is the study of the effect of massive neutrinos
(e.g. Lesgourges \& Pastor 2014, Abazajian et al. 2015).

Several studies of the nearby large-scale 
structure (LSS) of the Universe
have found indications of a discrepancy between their results
and the cosmological parameters obtained 
with {\sf Planck}, most importantly for
the cosmic matter density parameter, $\Omega_m$, and the amplitude
parameter of the matter density fluctuations on large 
scales, $\sigma_8$. These results
imply a smaller amplitude of the density fluctuations
and often a slightly smaller matter density than what is obtained,
if the structure characterised by {\sf Planck} is 
evolved forward to the present epoch
based on structure formation models for $\Lambda$CDM cosmology.
These LSS studies involve assessments of the 
cosmic lensing shear (e.g. Haman \& Hasenkamp 2013, Battye \& Moss 2014), 
redshift space distortions seen in galaxy
redshift surveys (e.g. Beutler et al. 2014, Ruiz \& Huterer 2015),
and the lensing of the CMB (Planck Collaboration 2015b). Most pronounced
is the discrepancy for galaxy clusters (e.g. Burenin 2012,
Planck Collaboration 2014, 2015c, B\"ohringer et al. 2014,
B\"ohringer \& Chon 2015).  

The discrepancy of the {\sf Planck} results and measurements 
of the present epoch LSS can in principle be reconciled by introducing
massive neutrinos as a small fraction of Dark Matter. 
We know from neutrino oscillations observed
for solar neutrinos, atmospheric
neutrinos and in terrestrial experiments that neutrinos have
mass, with a current lower limit for the sum of all 
three known neutrino families of $M_\nu = \sum~ m_{\nu_i} 
= 0.058$ eV (e.g. Lesgourges \& Pastor 2014). The measurement
of a useful upper
limit from terrestrial experiments is currently out of reach,
and we only have an upper limit for the mass of the electron 
neutrino of about 2 eV (e.g. Kraus et al. 2005). 
The study of the cosmic LSS already provides more stringent 
upper limits than this for the sum of all neutrinos,
since neutrinos with such high masses
would wash out too much of the LSS of the 
Universe (e.g. Riemer-S\o rensen et al. 2013). 
A careful comparison of the LSS imprinted on the CMB at 
an early epoch with the LSS measured in the nearby Universe
therefore opens thus
the very exciting opportunity to obtain constraints on neutrino 
masses from astronomical observations. 

In this paper we describe one of the most significant
of these results obtained with our galaxy cluster redshift survey
{\sf REFLEX} (B\"ohringer et al. 2013, 2014), discuss the systematics
of the results for galaxy clusters, and compare to other cosmological 
studies.

\section{The effect of massive neutrinos 
on cosmic structure evolution}

Massive neutrinos, as part of the Dark Matter in the Universe, 
have a clear effect on cosmological structure formation.
The later the dark matter becomes non-relativistic, the
more the smaller scales of Dark Matter fluctuations
will be damped or washed out. The fact that relavitistic 
Dark Matter particles can freely stream at the speed of light
causes density fluctuations in this relativistic
matter component to be erased on horizon scale. 
Therefore the main component
of the Dark Matter has to be ``cold'', that is, has
to become non-relativistic very early when the comoving horizon
size is smaller than galaxy scales, otherwise the galaxy population
would look very different from what we observe. But a smaller
admixture of hot Dark Matter, to which neutrinos are counted,
would have until recently remained undetected by observations.

\vspace{-0.4cm}
\begin{figure}[h]
\hbox{
\includegraphics[width=2.55in]{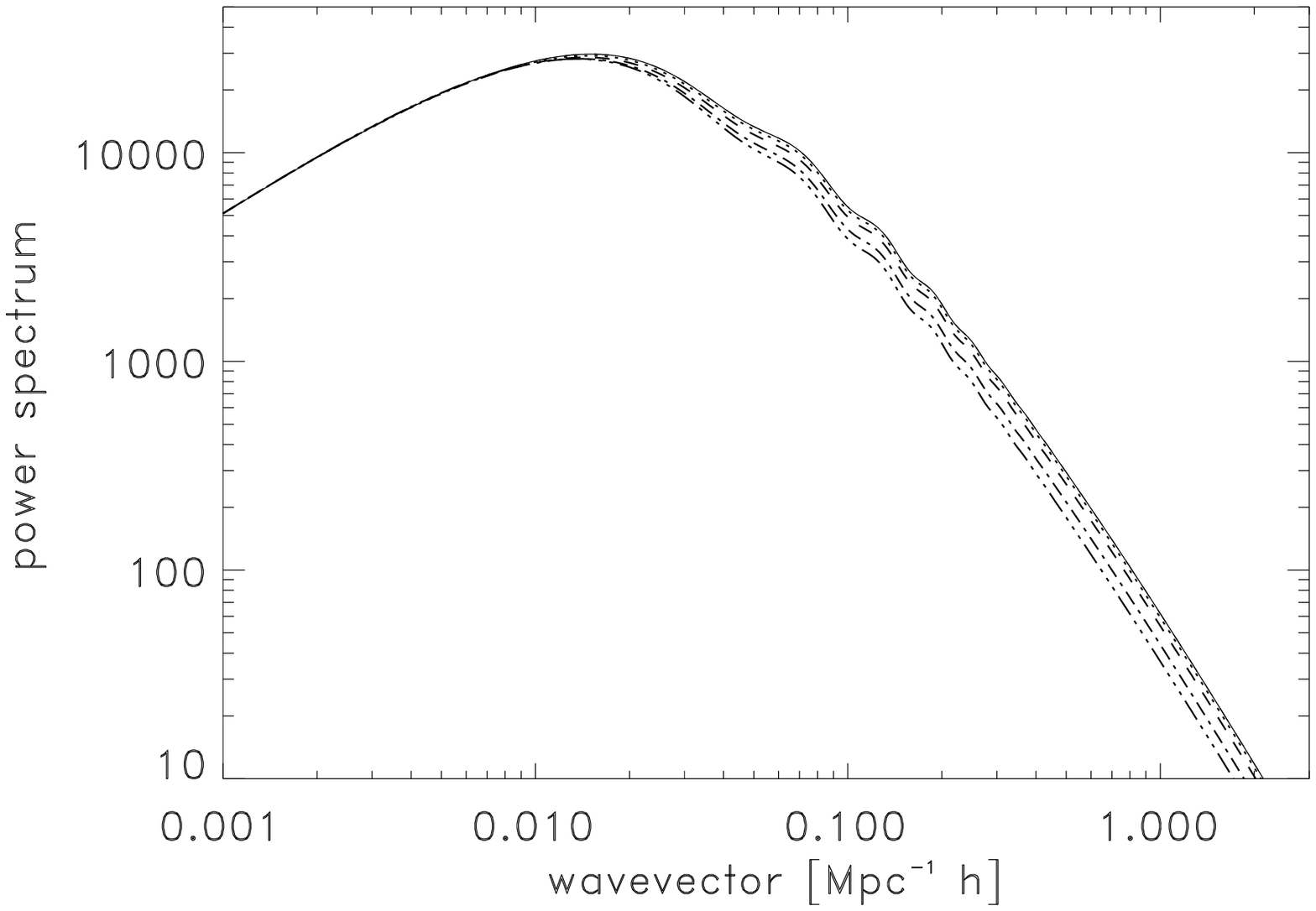}
\hspace{0.0cm}
\includegraphics[width=2.55in]{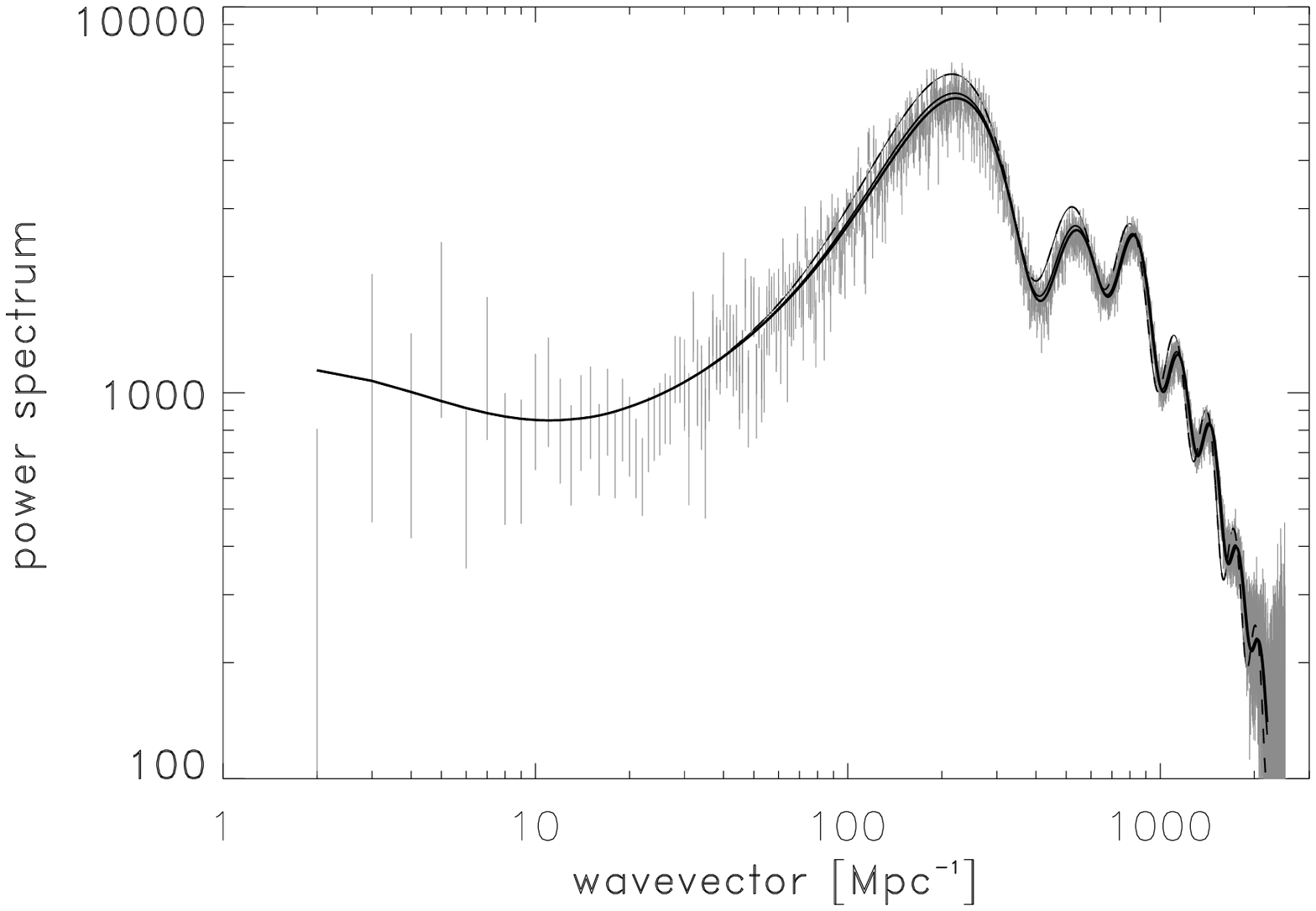}}
\caption{{\bf Left:} Effect of massive neutrinos on the present 
epoch matter density power spectrum. The solid curve shows the 
power spectrum for the sum of the neutrino masses, $M_{\nu} = 0.06$ eV, 
and the dashed lines show the cases for  $M_{\nu} = 0.17, 0.4, 0.6$ eV, 
from top to bottom. 
{\bf Right:} CMB power spectrum observed with {\sf Planck} 
and model predictions
for the best fitting cosmological parameters from the {\sf Planck}
survey for  $M_{\nu} = 0$ and $0.6$ eV (lower and upper solid line,
respectively). The dashed line artificially amplifies the difference 
between the two solid lines by a factor of 5.}  
\protect\label{fig1}
\end{figure}

\vspace{-0.3cm}
Neutrinos with a mass around 0.3 eV become non-relativistic
at redshifts around 600 (e.g. Abazajian et al. 2015). 
Thus density structure in
the neutrino matter component will be washed out on horizon
scales well after recombination, which will affect
the scales important for galaxy cluster formation. This effect 
on the matter power spectrum is shown in Fig. 1. 
The more massive the 
neutrinos are, the larger the dark matter fraction that is washed
out, since the number density of cosmic neutrinos is fixed
by the thermodynamics in the early Universe.

The effect of the neutrinos on the CMB power spectrum
is much smaller, as shown in the right panel of
Fig. 1. Therefore having a precise measurement of the 
amplitude of the present day matter power spectrum in
comparison to the CMB spectrum provides a means to
constrain the sum of all neutrino masses. 
The galaxy cluster mass function is exponentialy 
sensitive to the amplitude
of the matter power spectrum on scales of the order of 10 Mpc.
Therefore galaxy clusters provide an ideal tool to measure
this effect of neutrinos if combined with precise measurements
of the CMB.

\section{Probing cosmological parameters and neutrino
masses with the REFLEX galaxy cluster sample}

For the construction of the cluster mass function, a statistically 
complete survey of galaxy clusters with known masses is required.
One of the best ways to obtain such a sample remains an X-ray survey,
since X-ray emission of the hot ICM (few $\times$ 10$^7$ K 
plasma) is an unambiguous indication of a massive gravitational
potential and the X-ray luminosity is closely related to the total
mass of a clusters (e.g. Pratt et al. 2009, B\"ohringer et al. 2012).   
With the main goal to conduct cosmological studies of this kind, 
we have been performing large X-ray galaxy cluster surveys 
based on the ROSAT All-Sky Survey (Tr\"umper 1993, Voges et al. 1999)
with an extensive spectroscopic follow-up program
(B\"ohringer et al. 2000, 2001, 2004, 2013, Chon \& B\"ohringer 2012).
The {\sf REFLEX II} galaxy cluster survey covers
the southern sky below equatorial latitude +2.5$^o$ and at galactic
latitude $|b_{II}| \ge 20^o$, excluding the Magellanic Cloud regions
and comprises a survey area of $ \sim 4.24$ ster. It yielded a catalogue 
of 911 clusters. The nominal flux-limit is
$1.8 \times 10^{-12}$ erg s$^{-1}$ cm$^{-2}$ in the
0.1 - 2.4 keV energy band, which is reached in 78\% of the sky. The regions
where the flux limit is slightly higher are accounted for in the 
selection function. The source detection and characterisation, the
galaxy cluster sample definition and compilation, and the construction of
the survey selection function  as well as tests of the completeness of the
survey are described in B\"ohringer et al. (2013). 

The X-ray luminosity of clusters is 
determined within a radius if $r_{500}$\footnote{$r_{500}$ 
is the radius where the average
mass density inside reaches 500 times the critical density
of the Universe at the epoch of observation.}. 
The cluster masses are estimated from the X-ray luminosities
with the scaling relation (B\"ohringer et al. 2013),
$M_{500}~ =~2.48~ L_{500}^{0.62} ~E(z)^{-1}~ h_{70}^{-0.242}$,
where $M_{500}$ is the mass inside a radius of $r_{500}$, 
$E(z) = (\Omega_m (1+z)^3 + \Omega_{\Lambda})^{1/2}$, and $h_{70}$ is
the Hubble constant in units of 70 km s$^{-1}$ Mpc$^{-1}$. This relation
was derived in a detailed study of a representative subsample
of REFLEX galaxy clusters, REXCESS (B\"ohringer et al. 2007,
Pratt et al. 2009). From this cluster sample with 
the well known survey selection function we can construct the X-ray 
luminosity function with high precision.

\begin{figure}
\centerline{\includegraphics[width=3.5in]{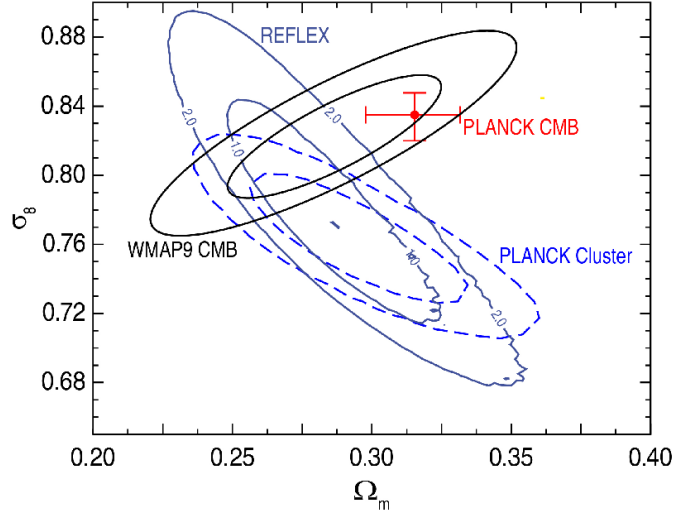}}
\caption{Constraints on $\Omega_m$ and $\sigma_8$
from the {\sf REFLEX II} X-ray luminosity function. 
The contours give the 1 and $2\sigma$ confidence intervals.
We compare this result to the constraints from the
galaxy cluster sample from {\sf Planck} and the constraints from
the {\sf Planck} and {\sf WMAP} CMB analysis.
}\label{fig2}
\end{figure}

To obtain cosmological model parameter constraints, 
the theoretical predictions for the cluster 
X-ray luminosity function are compared 
with the observations from the {\sf REFLEX II} project, as
described in B\"ohringer et al. 2014. The prediction
of the X-ray luminosity function involves the following steps:
First the power spectrum of the matter density
fluctuations for the present epoch is calculated with  
CAMB (Lewis et al. 2000)
\footnote{ CAMB is publicly available from
http://www.camb.info}. 
Based on this power spectrum we calculated the cluster 
mass function with the formulas
given by Tinker et al. (2008). 
From the mass function the X-ray luminosity function is 
derived using an empirical X-ray luminosity--mass relation
given above, based on the results from the {\sf REXCESS} survey 
(Pratt et al. 2009), in good agreement with Vikhlinin 
et al. (2009) and Reiprich \& B\"ohringer (2002).

In determining best fit model parameters through a 
likelihood analysis, we marginalised over an uncertainty
of the scaling relation slope of 7\% and normalisation
of 14\%. A scatter of the relation of 30\% and a typical 
measurement uncertainty for the X-ray luminosity of 20\%
were taken into account. 

Galaxy clusters are best at providing
constraints on the cosmological parameters, $\Omega_m$,
and $\sigma_8$. The constraints for these parameters are shown
in Fig. 2, where we compare our results to the 
constraints from an analysis of the galaxy cluster population detected
in the {\sf Planck} Survey (Planck Collaboration 2015b). 
There is excellent agreement between the two
results. The {\sf Planck} cluster sample is based on very different
selection criteria and covers a larger redshift range, providing
confirmation for the selection strategy in both surveys.
We also compare these results to the constraints derived
from the CMB anisotropies with {\sf WMAP} 
(Hinshaw et al. 2013) and {\sf Planck} (Planck Collaboration 2015a).
With the improved precision of the {\sf Planck} results, a discrepancy
between the cluster and the CMB parameter constraints now becomes apparent. 
Note that the comparison is made with observations of the
Universe at very different epochs. Thus the comparison is model dependent
and has only been performed here for a sum of the neutrino masses
of $0.06$ eV (Planck Collaboration 2015a). It has to be seen
if the discrepancy can be reconciled with a different cosmological model
(e.g. including more massive neutrinos).

\section{Constraints on the mass of the three neutrino families}

\begin{figure}
\hbox{
\hspace{-0.7cm}
\includegraphics[width=2.70in]{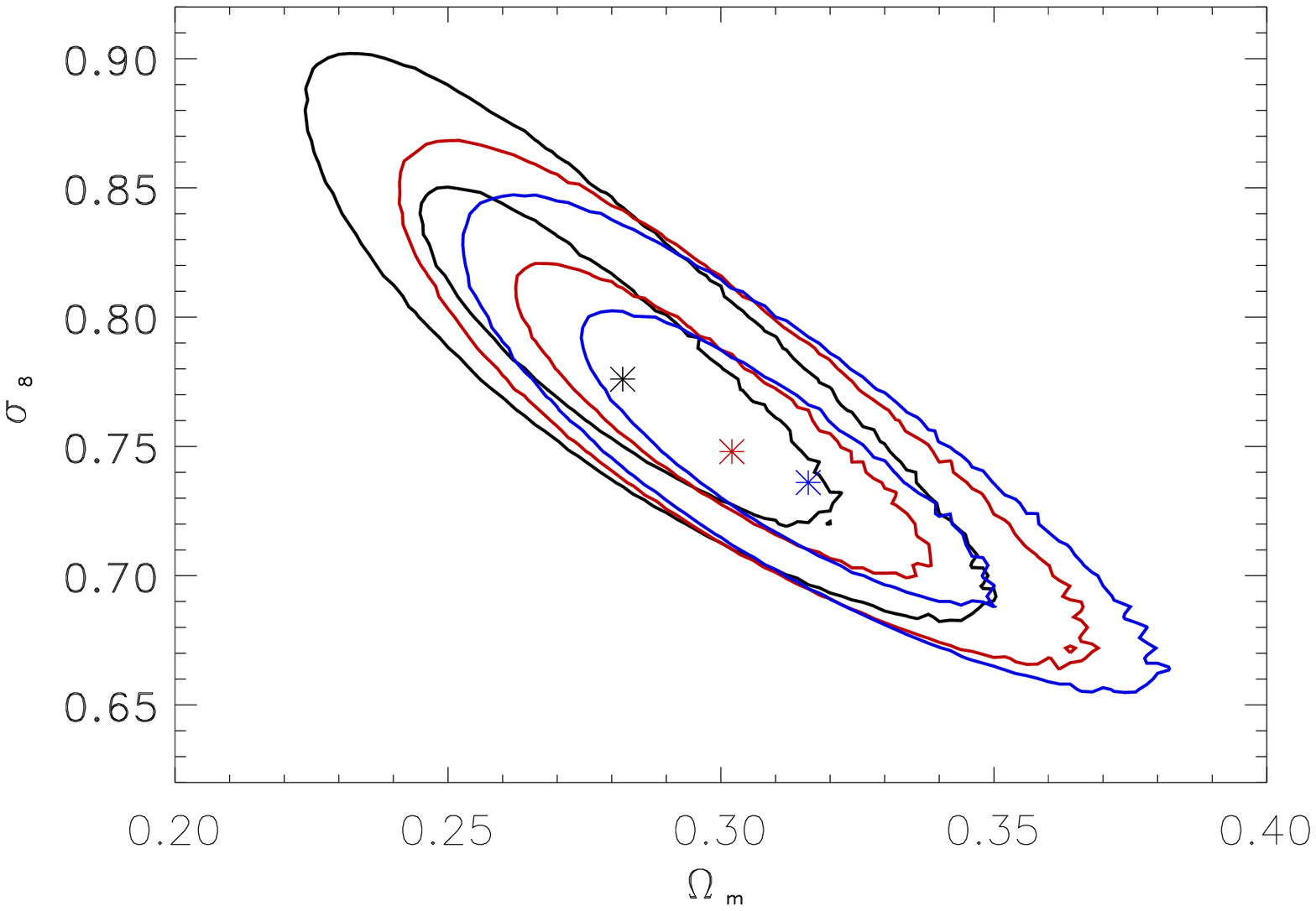}
\hspace{0.0cm}
\includegraphics[width=2.70in]{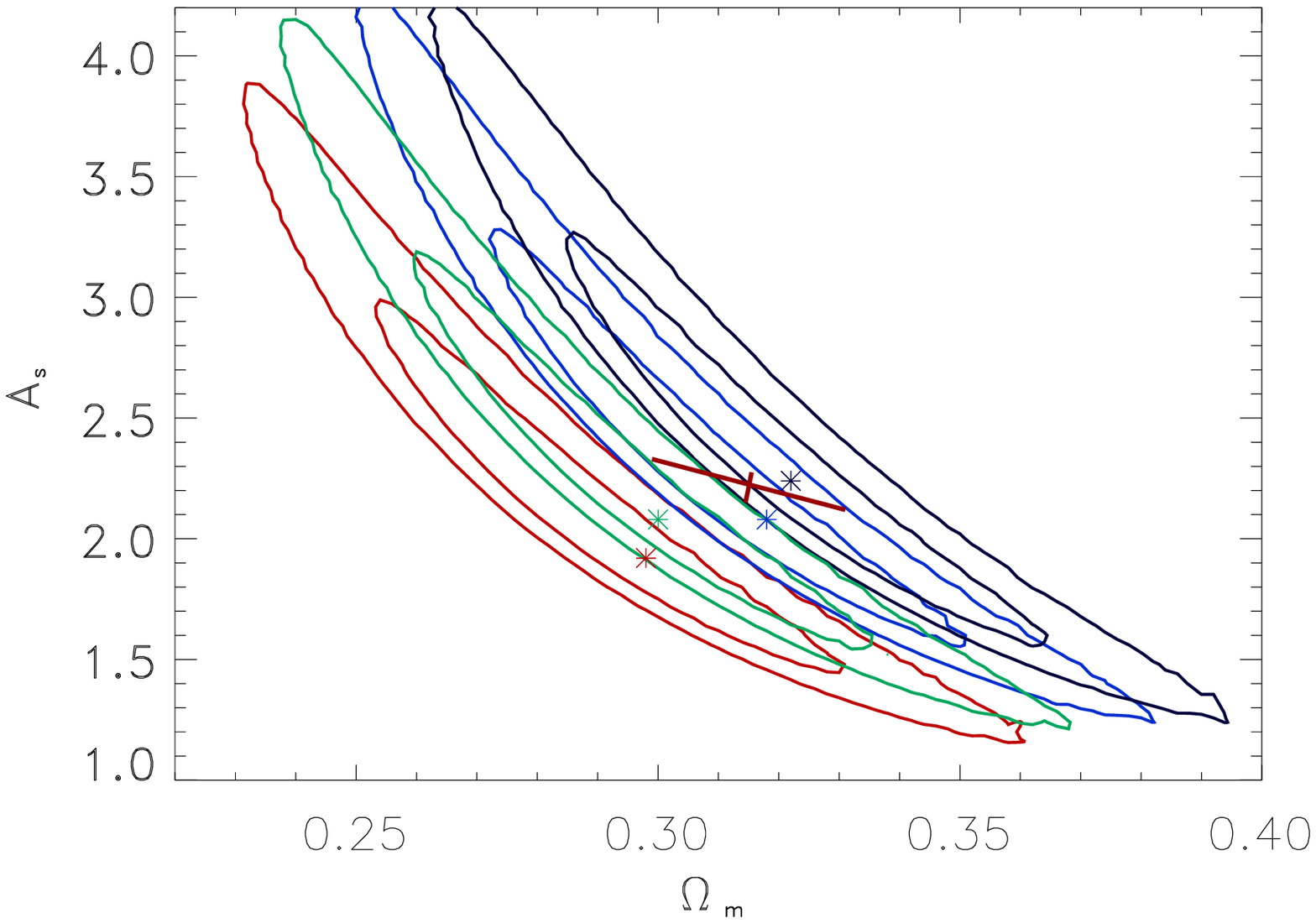}}
\vspace*{8pt}
\caption{{\bf Left:} Constraints on $\Omega_m$ and $\sigma_8$
from the {\sf REFLEX II} X-ray luminosity function 
for $M_{\nu} = 0$, 0.17, 0.4, and 0.6 eV, for the contours
from bottom to top, respectively. The contours give the 1 and
$2\sigma$ confidence intervals.
{\bf Right:} Constraints on $A_s$ and $\sigma_8$
from the {\sf REFLEX II} X-ray luminosity function for
$M_{\nu} = 0$, 0.17, 0.4, and 0.6 eV, for the contours
from bottom to top, respectively. The contours give the 1 and
$2\sigma$ confidence intervals. We also show the
constraints derived from the {\sf Planck} CMB observations
(data point with $1\sigma$ errors bars).
}\label{fig4}
\end{figure}

In the left panel of Fig. 3 we show the constraints on $\Omega_m$ 
and $\sigma_8$ obtained from the {\sf REFLEX} sample
when we allow the neutrinos to have masses in the range 
$M_{\nu} = 0 - 0.6$ eV. Since  $\sigma_8$ normalises the matter
power spectrum at a scale of $8 h^{-1}$ Mpc, the effect of
the neutrinos is almost normalised out. To reveal the effect
more clearly
and to better compare with the {\sf Planck} results,
we chose a different normalisation of the power spectrum, 
taken at large scales at the time of 
recombination with the parameter $A_S$,
also used for {\sf Planck}. The result is 
shown in the right panel of Fig. 3. Due to the small change
of the CMB power spectrum (as shown in Fig. 1), the
CMB constraints hardly change with varying neutrino mass
and we show only
one data point for the {\sf Planck} CMB.
For the REFLEX data we show the constraints for
four different total neutrino masses, $M_{\nu} = 0, 0.17, 0.4, 
0.6$ eV. We note that the constraints are in agreement for
neutrino masses in the range $M_{\nu} = 0.45 \pm 0.28$ eV
(1 $\sigma$ confidence, B\"ohringer \& Chon 2015). It is quite 
striking how the lightest particles in the Universe with masses 
of few $10^{-37}$ kg can influence the most massive systems,
galaxy clusters, with masses of the order of $10^{48}$ kg.

\section{Systematics}

The precision of the current results of cosmological studies
with galaxy clusters is limited by systematics, mostly by the
accuracy of the cluster mass estimates. The present study 
rests on the calibration of the  X-ray luminosity--mass relation
with deep X-ray studies, for which hydrostatic equilibrium
and spherical symmetry of the clusters were assumed.  
We know from detailed numerical hydrodynamical studies, that
this mass measurement is biased low by about 10 - 25\% because 
of the neglect of dynamical pressure of the ICM.
In our analysis we have assumed
that the hydrostatic masses are biased low by 10\% and 
adopted an additional uncertainty of 14\%.   

\begin{figure}
\hbox{
\includegraphics[width=3cm]{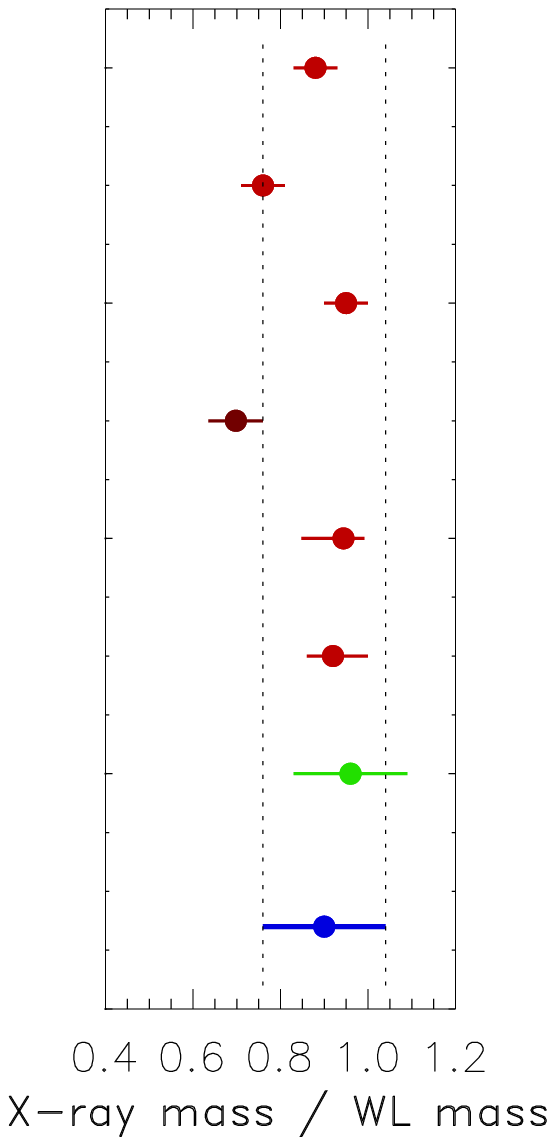}
\hspace{2.0cm}
\includegraphics[width=7cm]{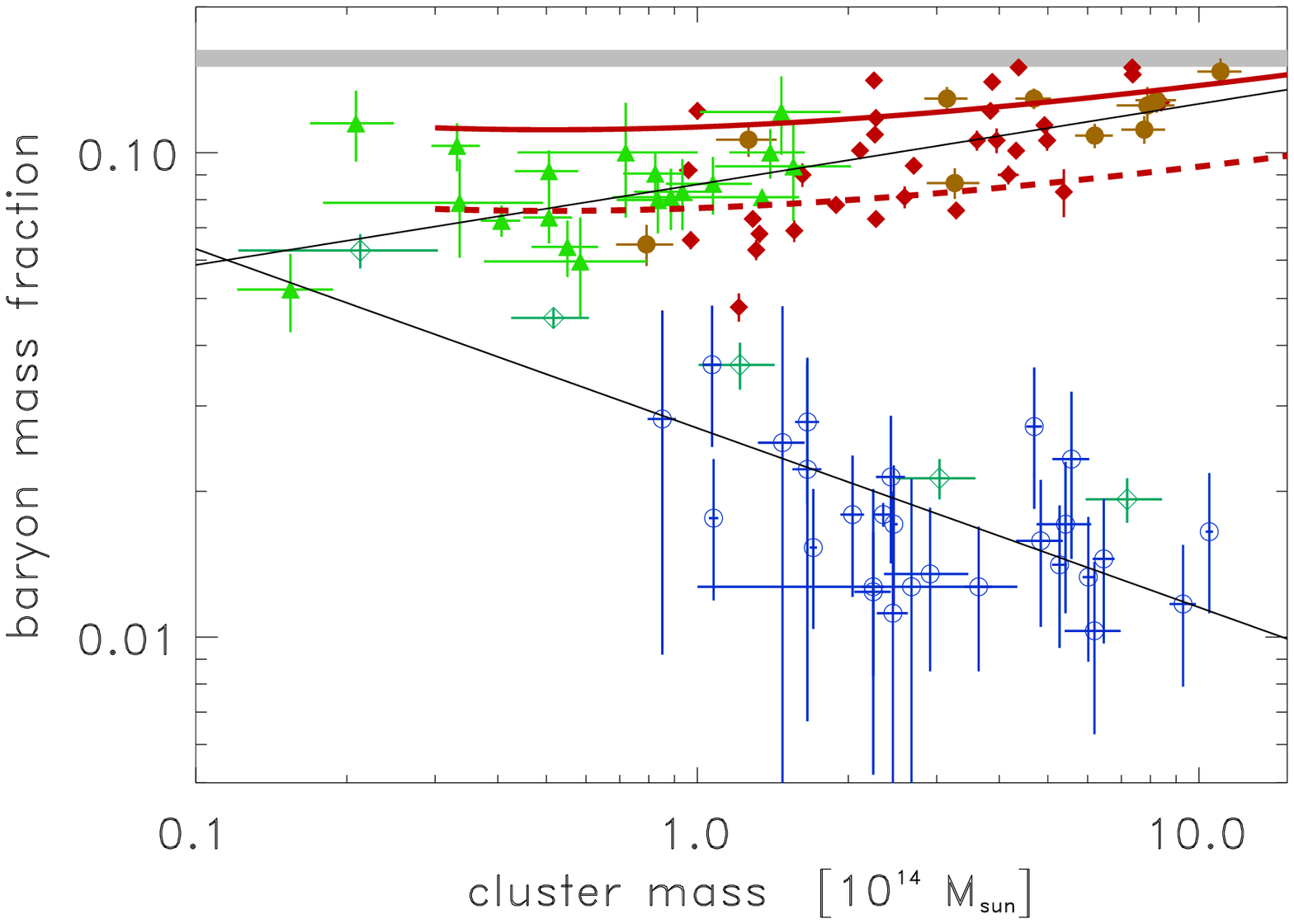}}
\caption{{\bf Left:} Ratios of cluster masses determined by
X-ray observations to those from gravitational lensing studies 
and their uncertainties for several cluster samples.
The lowest data point gives the mass bias and its
uncertainty assumed in our analysis of the REFLEX sample. 
{\bf Right:} Gas (filled symbols) and stellar mass fractions (open
symbols) from Vikhlinin et al. 2006 (circles), Pratt et al. 2009
(diamonds), Sun et al. 2009 (triangles), Lin et al 2003 (open
circles), and Giodini et al. 2009 (binned data, open diamonds).
The solid lines are (from bottom to top), fitted power laws to
the stellar and gas mass fractions and the total baryon fraction
as the sum of the two. The grey shaded area shows 
the cosmic baryon fraction 
(Planck Collaboration 2015a) and the dashed line indicates
the value for the cluster baryon fraction if a mass bias
of 32\% would have been adopted. 
}\label{fig6}
\end{figure}

Another source of information on the possible mass bias comes
from the comparison of X-ray determined
hydrostatic masses with cluster masses derived from weak
gravitational lensing analyses. While the latter measurement
has a substantial individual scatter, it is believed to have
a very small bias (e.g. Becker \& Kravtsov 2003). Recent results
of the comparison of X-ray or Sunyaev-Zeldovich effect based
mass measurements with weak lensing studies have produced 
results ranging from almost no bias to bias
values as large as 32\% (e.g. Mahdavi et al. 2013, 
Hoeckstra et al. 2015, Smith et al. 2016), where the bias
value here indicates the mass underestimate by the X-ray analysis.
If the bias would be larger than
30\%, the significance of the neutrino mass signal is lost
(e.g. Planck Collaboration 2015c, Mantz et al. 2015).
In the left panel of Fig. 4 we compare the range of the
mass bias ratios from the lensing results to our 
marginalisation, and apart from one outlier data point, our
marginalisation covers the observational results for the
mass calibration very well.

There is a good argument to rule out extreme values of
hydrostatic mass bias. The gas mass of the intracluster 
medium can be determined more easily than the total 
mass of a cluster. Since galaxy clusters 
are assembling cosmic matter almost indiscriminantly into their
gravitational potential, we expect an almost exact cosmic ratio
of the mass in baryons to that of the dark matter. In Fig. 4
we compare the fraction of the total mass contained in gas
and in stars to the total matter. We also compare the
summed baryon mass fraction with the cosmic ratio
determined with {\sf Planck}. We see that the cluster
baryon fraction falls short of the cosmic ratio, but less so for 
the most massive clusters. The reason for
the discrepancy is extra heat which is deposited in the gas during
galaxy and active galactic nuclei evolution, which inflates the gas
halo and pushes some of the baryonic matter beyond the fiducial 
radius of a cluster. This has little effect on the most massive 
clusters with deep gravitational potentials. 
For the data shown in Fig. 4
the total cluster masses have been determined assuming hydrostatic 
equilibrium. A small mass bias correction could be tolerated
in this study. But a mass bias as large as 30\% would imply that
almost half the baryons, mostly in the gas, have to be pushed out
of the cluster potentials even in the most massive clusters. 
This is extremely unlikely given
the large energy input needed and based on the experience
with hydrodynamic simulations of cluster formation.

\section{Comparison to other large-scale structure 
measurements}

The effect of massive neutrinos is best revealed through the 
comparison of the density fluctuation amplitude 
of the present epoch LSS and the prediction
of the $\Lambda$CDM model based on the  {\sf Planck} results.
Therefore it is interesting to check if other nearby LSS assessments,
like LSS gravitational lensing, the galaxy clustering power 
spectrum amplitude, redshift space distortion effects in the
galaxy correlation function, CMB lensing, and the study of the
structure of the Ly$\alpha$ forest show a similar trend requiring
massive neutrinos. This has been found in some of these studies,
which in particular involve the study of lensing and redshift 
space distortions as well as  clusters:
Beutler et al. (2014) found $M_{\nu} = 0.36 \pm 0.14$ eV (1$\sigma$)
from the SDSS power spectrum and redshift space distortion effects
in comparison with the CMB, Costanzi et al. (2014) obtain a result
of $M_{\nu} = 0.29 {+0.18 \atop -0.21}$ eV comparing lensing
and cluster observations to CMB results, also Ruiz \& Huterer
(2015), Battye \& Moss (2013), Hamann \& Hasenkamp (2013),
and Wyman et al. (2014) find evidence for
a non-zero neutrino mass based on lensing and/or cluster observations.
All these results are in good agreement with our results from
the REFLEX cluster survey. Analyses of just the galaxy power 
spectrum or the power spectrum of the Ly$\alpha$ forest have 
recently produced only upper limits, which are quite
tight and only marginally consistent with our results. Among
those are the study of Riemer-S\o rensen et al. (2014) and
Cuesta et al. (2015) who combined the SDSS and WiggleZ galaxy
power spectrum with BAO signatures and the {\sf Planck} results
finding $M_{\nu} \le 0.18$ eV and $M_{\nu} \le 0.12$ eV,
respectively. Also Ly$\alpha$ studies yielded lower upper
limits, of e.g.  $M_{\nu} \le 0.17$ (Seljak et al. 2006) and
$M_{\nu} \le 0.12$ (Palanque-Dellabrioulle et al. 2015).
Massive neutrinos also have a subtle effect the CMB
measurable with {\sf Planck} yielding $M_{\nu} \le 0.49$ eV. 
Combined with BAO observations (which are not sensitive
to LSS amplitudes, but help to break degeneracies in the set of 
cosmological parameters) the limit is $M_{\nu} \le 0.21$ eV
(Planck Collaboration 2015a). Lensing in the CMB provides 
another measure of the low redshift LSS.
The analysis of CMB lensing in the {\sf Planck} Survey  
(Planck Collaboration 2015b) shows indeed a trend that 
the lensing signal and thus the local LSS amplitude is 
smaller than expected from the CMB results in a $\Lambda$CDM
model. Thus even so there are several indications 
that neutrinos have more than the minimum mass of
$M_{\nu} = 0.06$, the overall results are not in
perfect agreement.
  
\section{Summary and Conclusion}

Cosmological studies with {\sf Planck} together with LSS observations
have clearly shown that a background of cosmic neutrinos 
exists, providing a constraint
for the effective number of neutrino species 
of $N_{eff} = 3.15 \pm 0.23$ (Planck Collaboration 2015a).
With the particle physics lower limit for the sum of the 
neutrino masses of $M_{\nu} = 0.058$ eV, neutrinos 
contribute at least about 0.4\% to the total matter
content in the Universe and have a significant effect
on the evolution of the cosmic large-scale structure.
This effect is best revealed by a comparison of the
{\sf Planck} CMB study with present epoch LSS measurements.

Several such studies have provided evidence for a non-zero
neutrino mass with galaxy clusters showing one of the
strongest signals. We discussed here the results
from the REFLEX cluster survey with the largest well-defined
X-ray cluster sample for this type of
study, yielding tight constraints 
on  the cosmological parameters  $\sigma_{8}$ and $\Omega_m$.
A comparison with the results from {\sf Planck} suggests
a sum of the neutrino masses in the range 
$M_{\nu} = 0.17 - 0.73$ eV (1$\sigma$). Other LSS measurements 
which show a similar trend include gravitational lensing
and redshift space distortion observations, while the study
of the galaxy power spectrum and the structure
of the Ly$\alpha$ forest yield smaller upper limits which are
only marginally consistent with the positive neutrino mass
detections. 

We have in this paper not considered sterile neutrinos. 
A main motivation for the inclusion of sterile neutrinos
in cosmological models was the discrepancy in the Hubble
constant determined locally (inside a radius smaller than 200 Mpc)
and with {\sf Planck}. We have recently shown, based on
the distribution of galaxy clusters, that we live in
a locally underdense region in the Universe, at least for
the southern sky region (B\"ohringer et al. 2015), which 
can at least resolve a part of the tension between the local
Hubble constant measurements and {\sf Planck}.  

All these studies are systematics limited. For clusters the
most serious limiting systematic is the cluster mass calibration.
Large efforts are currently underway to beat down these uncertainties,
to fully exploit the potential of a cosmological neutrino
mass measurement. Several future surveys include constraining
of the neutrino mass as an important science goal, for example,
the sky surveys with DESI, LSST and EUCLID and possibly a
new advanced CMB mission. The aim for these projects is an
uncertainty of $\Delta M_{\nu} \sim 0.025 (0.016)$ eV and
$N_{eff} \sim 0.08 (0.02)$, where the numbers in brackets refer
to results including an advanced CMB mission (e.g. Abazajian
et al. 2015). With this outlook we can hope that the currently 
more qualitative results can turn into a fundamental precision 
measurement for particle physics.

\section*{Acknowledgments}

H.B. and G.C. acknowledge support from the DFG Transregio Program TR33
and the Munich Excellence Cluster ``Structure and Evolution of the Universe''.  
G.C. acknowledges support by the DLR under grant no. 50 OR 1405.


\end{document}